\documentclass[twocolumn,floatfix,aps,prb]{revtex4}
\usepackage{graphicx}
\usepackage{amsmath}
\usepackage{amssymb}
\usepackage{bm}
\usepackage{hyperref}

\usepackage{color}

\begin{document}

\title{Gapless odd-frequency superconductivity induced by the Sachdev-Ye-Kitaev model}
\author{Nikolay V. Gnezdilov}
\email{gnezdilov@lorentz.leidenuniv.nl}
\affiliation{Instituut-Lorentz, Universiteit Leiden, P.O. Box 9506, 2300 RA Leiden, The Netherlands}

\begin{abstract}
We show that a single fermion quantum dot acquires odd-frequency  Gor'kov anomalous averages in proximity to strongly-correlated Majorana zero-modes, described by the Sachdev-Ye-Kitaev (SYK) model. Despite the presence of finite anomalous pairing, superconducting gap vanishes for the intermediate coupling strength between the quantum dot and Majoranas. The increase of the coupling leads to smooth suppression of the original quasiparticles. 
This effect might be used as a characterization tool for recently proposed tabletop realizations of the SYK model.
%The spectral function of the quantum dot in this regime is qualitatively similar to one observed in the tunneling spectroscopy of high-$\mathrm{T_c}$ superconductors with pseudogap. The increase of the coupling leads to smooth suppression of the particle-hole mixing and restoration of quasiparticles. 
\end{abstract}

\date{October 2018}

\maketitle

\textit{Introduction ---} 
The Sachdev-Ye-Kitaev (SYK) model \cite{SY, Kitaev} describes $N$ fermionic zero-modes with randomized infinite-range interaction. It comprises several important properties: (i) the SYK model possesses an exact large $N$ solution in the infrared lacking  quasiparticles; (ii) it saturates \cite{Kitaev, Maldacena1} the upper bound on quantum chaos \cite{Maldacena2}, which is also the case for holographic duals of black hole horizons \cite{Hartnoll}. 
A possibility to study these intriguing properties in physical observables  inspired a few proposals of realizing the SYK model in a solid-state platform \cite{Pikulin, Alicea, Pikulin2}. 

The SYK model with Majorana (real) zero-modes is claimed to be a low-energy theory of the Fu-Kane superconductor \cite{FK}  in a magnetic field with a disordered opening\cite{Pikulin}, whereas Ref.\,\onlinecite{Alicea} suggests to use $N$ Majorana nanowires \cite{DasSarma} coupled through a disordered quantum dot. The graphene flake device proposed in Ref.\,\onlinecite{Pikulin2} realizes the SYK model with the conventional (complex) fermionic zero-modes (cSYK model)\cite{Sachdev_BH}. As for the latter one, the signatures of non-Fermi liquid/non-quasiparticle/quantum critical behavior \cite{Sachdev, Hartnoll} of the cSYK model have been recently studied in Refs.\,\onlinecite{dIdV_SYK, Franz, Zeros}. The one dimensional extensions of the cSYK model to the coupled clusters uncover the Lyapunov time $\vert$the characteristic timescale of quantum chaos$\vert$ in thermal diffusion \cite{Sachdev_1D} and demonstrate linear in temperature resistivity of strange metals \cite{Balents}. 

In this paper, we modify the SYK model with Majoranas via coupling it to a single-state non-interacting quantum dot. As we add only a single fermion, this model stays far away from the non-Fermi liquid/Fermi liquid transition \cite{Altman} and it is still exactly solvable in the large $N$ limit. We demonstrate that the effective theory for the fermion in the quantum dot gains the anomalous pairing terms, that make the quantum dot superconducting. Despite the induced superconductivity, the density of states in the quantum dot has no excitation gap. It has been a while since the phenomenon of gapless superconductivity was found in the superconductors with magnetic impurities, where for a specific range of concentration of those, a part of electrons does not participate in the condensation process \cite{Abrikosov, Woolf}. The anomalous components of the Gor'kov Green's function \cite{Gorkov, AS} of the quantum dot are calculated exactly in the large $N$ limit and are odd functions of frequency \cite{Berezinskii, Balatsky}. Odd-frequency pairing is known to be induced by proximity to an unconventional superconductor \cite{Balatsky, Lutchyn1, Tanaka, Tanaka_review}.
Below we obtain induced odd-frequency gapless superconductivity in zero dimensions as a consequence of the proximity to a system described by the SYK model\cite{Pikulin, Alicea}. We suggest to use this effect as a way to detect the SYK-like effective behavior in a solid-state system.

\textit{The model ---}   Let's consider the Sachdev-Ye-Kitaev model \cite{Kitaev,Maldacena1} randomly coupled to a single state quantum dot \cite{Kouwenhoven} with the frequency $\Omega_d$. The Hamiltonian of the system reads:
\begin{align}
\hspace{-0.5em}H\!\!=\!\Omega_d d^\dag d \!+ \!\!\sum_{i=1}^N \lambda_i \gamma_i \left(d^\dag\!\!-\!d\right)\!+\!\frac{1}{4!}\!\!\sum_{i,j,k,l=1}^N \!\! J_{ijkl} \gamma_i \gamma_j \gamma_k \gamma_l\, \label{H},
\end{align}
where the couplings $J_{ijkl}$ and $\lambda_i$ are independently distributed as a Gaussian with zero mean $\overline{J_{ijkl}}=0=\overline{\lambda_i}$ and finite variance $\overline{J_{ijkl}^2}=3! J^2/N^3$, \, $\overline{\lambda_i^2}=\lambda^2/N$. The tunneling term in the Hamiltonian (\ref{H}) is similar to one, that appears for tunneling into Majorana nanowires \cite{Majorana_review, Fisher, Lutchyn3, Lutchyn1, Lutchyn2}.

Once the disorder averaging is done, we decouple the interactions by introducing four pairs of the non-local fields in the Euclidean action as a resolution of unity\cite{Kitaev, Maldacena1}:
\begin{align}
1&{}=\!\!\int\!\! \mathcal{D} \Sigma_\gamma \mathcal{D} G_\gamma \mathrm{e}^{\int \!\! d \tau  d \tau' \frac{\Sigma_\gamma(\tau,\tau')}{2}\!\left(\!N \!G_\gamma(\tau',\tau)-\sum_i \!\!\gamma_i(\tau) \gamma_i(\tau')\!\right)} \label{1_gamma}, \\
1&{}=\!\!\int\!\! \mathcal{D} \Sigma_d \mathcal{D} G_d \mathrm{e}^{\int \!\! d \tau  d \tau' \Sigma_d(\tau,\tau') \left(\!G_d(\tau',\tau)-\bar{d}(\tau) d(\tau')\!\right)} \label{1_d*d}, \\
1&{}=\!\!\int\!\! \mathcal{D} \Xi_d \mathcal{D} F_d \mathrm{e}^{\int \!\! d \tau  d \tau' \Xi_d(\tau,\tau') \left(\!F_d(\tau',\tau)-d(\tau) d(\tau')\!\right)} \label{1_dd}, \\
1&{}=\!\!\int\!\! \mathcal{D} \bar{\Xi}_d \mathcal{D} \bar{F}_d \mathrm{e}^{\int \!\! d \tau  d \tau' \bar{\Xi}_d(\tau,\tau') \left(\!\bar{F}_d(\tau',\tau)-\bar{d}(\tau) \bar{d}(\tau')\!\right)} \label{1_d*d*}.
\end{align} 
A variation of the effective action, which is given in Appendix \ref{app:SC}, with respect to $G_\gamma, G_d, F_d, \bar{F}_d$ and  $\Sigma_\gamma, \Sigma_d, \Xi_d, \bar{\Xi}_d$  produces self-consistent Schwinger-Dyson equations \cite{AS}, that relate those fields to the Green's functions and self-energies of Majorana fermions and the fermion in the quantum dot:
\begin{align} 
\Sigma_d(\tau)=&{}\lambda^2 G_\gamma(\tau) \label{Sd_eq},\\ 
\Xi_d(\tau)=&{}-\frac{\lambda^2}{2} G_\gamma(\tau), \;\;\;
\bar{\Xi}_d(\tau)=-\frac{\lambda^2}{2} G_\gamma(\tau) \label{Xi_eq},\\ 
\Sigma_\gamma(\tau)=&{}J^2 G_\gamma(\tau)^3 +\!\frac{2\lambda^2}{N} \!\bigg(\!G_d(\tau)-\!\frac{F(\tau)}{2}\!-\!\frac{\bar{F}(\tau)}{2}\bigg) \label{S_eq}, \\
G_\gamma(\mathrm{i}\omega_n)=&{}\bigg(\mathrm{i}\omega_n-\Sigma_\gamma(\mathrm{i} \omega_n)\bigg)^{-1} \label{G_eq}.
\end{align}
The Green's function of Majorana fermions is $G_\gamma(\tau)=-N^{-1} \sum_i\left\langle \mathcal{T}_\tau\, \gamma_i(\tau) \gamma_i(0)\right\rangle$ and $G_d(\tau)=-\left\langle \mathcal{T}_\tau\, d(\tau) \bar{d}(0)\right\rangle$, $F_d(\tau)=-\left\langle \mathcal{T}_\tau\, d(\tau) d(0)\right\rangle$, $\bar{F}_d(\tau)=-\left\langle \mathcal{T}_\tau\, \bar{d}(\tau) \bar{d}(0)\right\rangle$ are normal and anomalous Green's functions of the quantum dot variables. 

We are focused on the large $N$, long time limit $1 \ll J \tau \ll N$, where the conformal symmetry of the SYK model emerges \cite{Kitaev, Maldacena1}. 
In this regime, the backreaction of the quantum dot on the self-energy of Majorana fermions (\ref{S_eq}) is suppressed as $1/N$. The bare frequency in the equation (\ref{G_eq}) can also be omitted at low frequencies. Thus, equations (\ref{S_eq}, \ref{G_eq}) become $\Sigma_\gamma(\tau)=J^2 G_\gamma(\tau)^3$ and $G_\gamma(\mathrm{i}\omega_n)=-\Sigma_\gamma(\mathrm{i} \omega_n)^{-1}$, which are the same as in the case of the isolated SYK model. 
These equations have a known zero temperature solution \cite{Kitaev, Maldacena1} $G_\gamma(\mathrm{i}\omega_n)=-\mathrm{i} \pi^{1/4} \mathrm{sgn}(\omega_n)\left(J| \omega_n|\right)^{-1/2}$, which contributes to the self-energies (\ref{Sd_eq}, \ref{Xi_eq}) of the quantum dot. The Green's function of Majorana zero-modes has no pole structure, which manifests the absence of quasiparticles. Moreover, it behaves as a power-law of frequency, which is the case of quantum criticality \cite{Sachdev} and emergence of the conformal symmetry in the SYK case \cite{Kitaev, Maldacena1}. 

\textit{SYK proximity effect ---} The effective action for the fermion in the quantum dot acquires anomalous terms
\begin{align}
S_{\rm eff}=-\frac{1}{2}\sum_{n=-\infty}^{+\infty} \begin{pmatrix}
\bar{d}_n & d_{-n}
\end{pmatrix} \mathcal{G}(\mathrm{i}\omega_n)^{-1} \begin{pmatrix}
d_n \\ \bar{d}_{-n}
\end{pmatrix} \label{S_eff},
\end{align}
where the Gor'kov Green's function \cite{Gorkov, AS} 
\begin{gather}
\hspace{-0.5em}\mathcal{G}(\mathrm{i} \omega_n\!)^{-1}\!=\!\!\begin{pmatrix}
\mathrm{i} \omega_n \!\!- \!\Omega_d\!-\!\lambda^2 G_\gamma(\mathrm{i}\omega_n\!)& \lambda^2 G_\gamma(\mathrm{i}\omega_n\!) \\ \lambda^2 G_\gamma(\mathrm{i}\omega_n\!) & \mathrm{i} \omega_n\!\!+\! \Omega_d\!-\!\lambda^2 G_\gamma(\mathrm{i}\omega_n\!)
\end{pmatrix} \label{GG} \!
\end{gather}
is found self-consistently in a one loop expansion \cite{AS}. Due to negligibility of the last term in Majoranas self-energy (\ref{S_eq}) mentioned above, the one loop approximation turns out to be exact in the large $N$ limit.  A detailed derivation of the formula (\ref{GG}) is presented in Appendix \ref{app:SC}.

\begin{figure}[ht!!]
\center
\includegraphics[width=1.\linewidth]{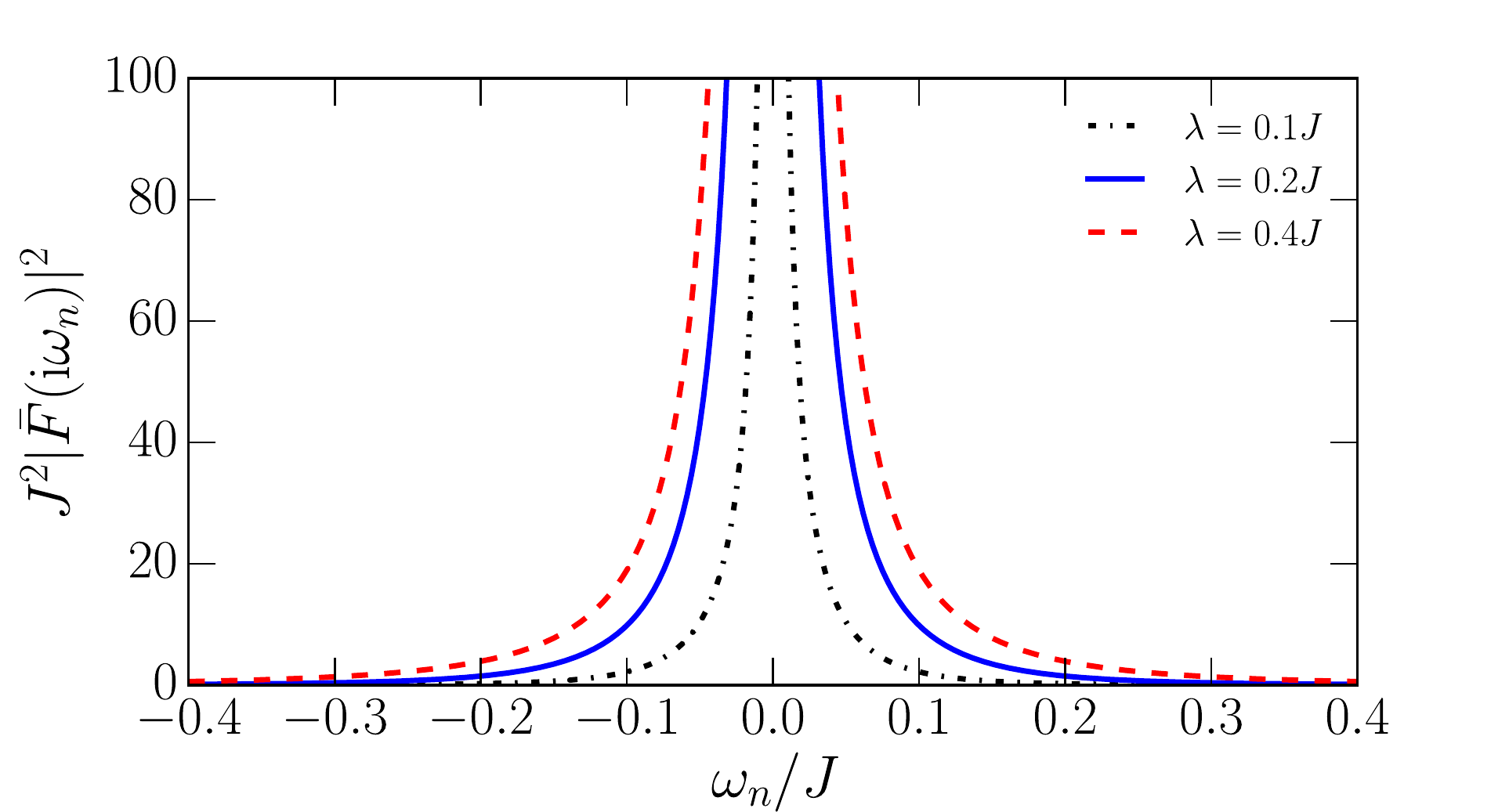}
\caption{\small \label{fig:FFdag} Absolute value of the \textbf{anomalous averages}  as a function of Matsubara frequency. The frequency of the quantum dot is $\Omega_d=0.1 J$. }
\end{figure}

Appearance of the anomalous pairing terms $\bar{d} (\tau) G_\gamma(\tau-\tau') \bar{d} (\tau')$ in the effective action (\ref{S_eff}) does not require any additional quantum numbers, because those are ``glued'' by the non-locality in the imaginary time that originates from the SYK saddle-point solution.
The anomalous Green's function which follows from  (\ref{GG}) is odd in frequency \cite{Berezinskii, Balatsky}:
\begin{align} \nonumber
\bar{F}(\mathrm{i} \omega_n) =&{}-\frac{\lambda^2 G_\gamma(\mathrm{i} \omega_n)}{\mathrm{i} \omega_n\left(\mathrm{i} \omega_n -2\lambda^2 G_\gamma(\mathrm{i} \omega_n)\right) -\Omega_d^2} =\\=&{}-\bar{F}(-\mathrm{i}\omega_n) \label{barF}.
\end{align}
This result (\ref{barF}) is well aligned with previously found proximity effect by Majorana zero modes \cite{Lutchyn1, Lutchyn2} and odd-frequency correlations found in interacting Majorana fermions \cite{Balatsky2}. Superconducting pairing grows smoothly while  the coupling increases as it is shown in FIG.\,\ref{fig:FFdag}.

\begin{figure}[ht!!]
\center
\includegraphics[width=1.\linewidth]{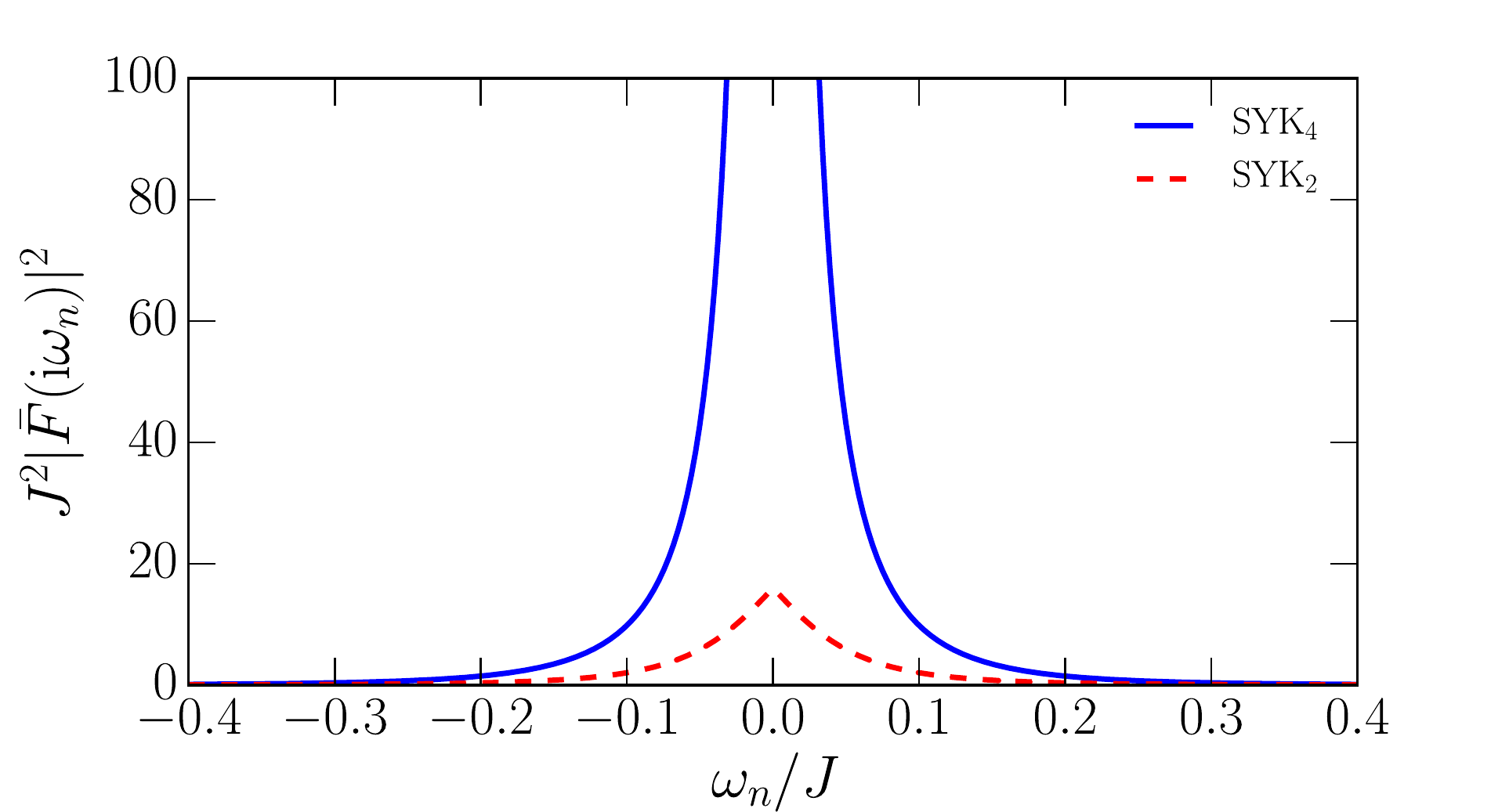}
\caption{\small \label{fig:FFdag24} \textbf{Anomalous averages} in the quantum dot coupled to the \textbf{SYK/SYK$_2$} model. The frequency of the dot is $\Omega_d=0.1 J$ and the coupling strength is $\lambda=0.2 J$.}
\end{figure}

It is worthwhile to compare our setting (\ref{H}) to the case when the SYK quantum dot is replaced by a disordered Fermi liquid. The latter can be described by the SYK$_2$ model: $H_{{\rm SYK}_2}=\mathrm{i} \sum_{ij}J_{ij}\gamma_i \gamma_j$. In the long time limit, the Green's function of the SYK$_2$ model is $G_{{\rm SYK}_2}(\mathrm{i} \omega_n)=-\mathrm{i} \, \mathrm{sgn}(\omega_n)/J$ \cite{Pikulin}, which is substituted to the result for the anomalous component of the Gor'kov  function (\ref{barF}).
As we show in FIG. \ref{fig:FFdag24}, the amount of the SYK induced  superconductivity is sufficiently higher then in the case of the SYK$_2$ model.

\begin{figure}[ht!!]
\center
\includegraphics[width=1.\linewidth]{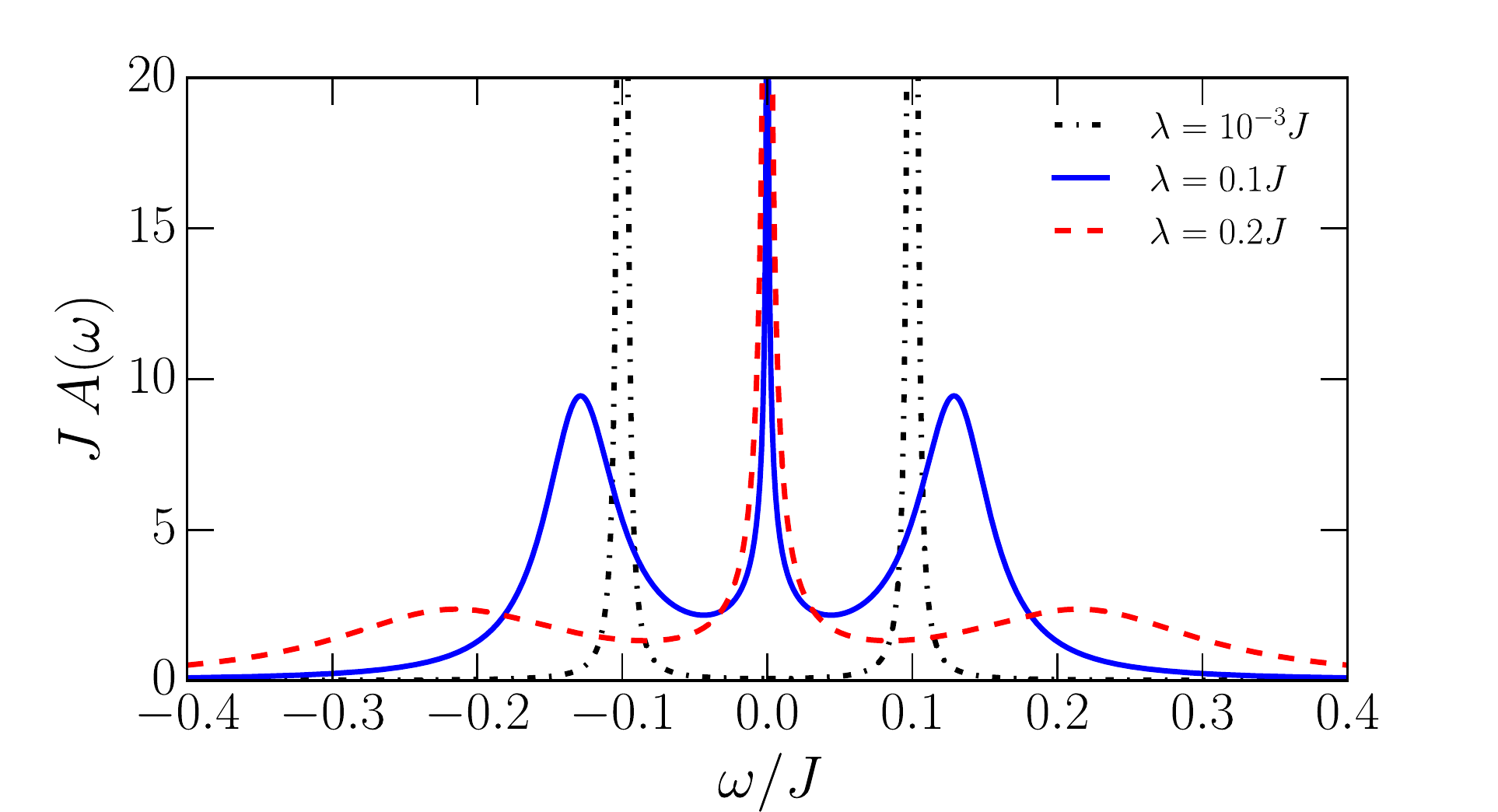}
\caption{\small \label{fig:LDOS} \textbf{Density of states in the quantum dot at zero temperature} as a function of frequency. $\Omega_d=0.1 J$ and $\delta=10^{-3} J$.}
\end{figure}

In the large $N$ limit, the spectral function of the quantum dot is 
\begin{align} \nonumber
A(\omega)=&{}-\frac{1}{\pi} \mathrm{Im} \, \mathrm{tr} \mathcal{G}\left(\mathrm{i}\omega_n \to \omega+\mathrm{i}\delta \right) = \\=&{} -\frac{2}{\pi}\,\frac{\lambda^2 \left(\omega^2+\Omega_d^2\right)\mathrm{Im}G^R_\gamma(\omega)
}{\left|\omega\left(\omega -2\lambda^2 G^R_\gamma(\omega)\right)-\Omega_d^2\right|^2}\, \label{LDOS},
\end{align}
where $\delta=0^+$ and $G_\gamma^R(\omega)=-\mathrm{i} \pi^{1/4} \mathrm{e}^{\mathrm{i} \pi \mathrm{sgn}(\omega)/4} \left(J|\omega|\right)^{-1/2}$.
The broadening $\delta=0^+$ of the fermion in the quantum dot is neglected once the imaginary part of the SYK Green's function is finite: $\lambda^2 \mathrm{Im} G^R_\gamma(\omega) \gg \delta =0^+$. 

In absence of coupling between the single-state quantum dot and the SYK model ($\lambda=0$), there is no particle-hole mixing. Superconducting pairing (FIG.\,\ref{fig:FFdag}) appears in the regime of intermediate coupling. 
The absence of the gap in the presence of the anomalous pairing reveals gapless superconductivity \cite{Abrikosov, Woolf} in zero dimensions, which can be probed by Andreev reflection \cite{Andreev} in the tunneling experiment.
The wide broadening of the peaks in FIG.\,\ref{fig:LDOS} is due to the binding of the fermionic quantum dot with the SYK quantum critical continuum \cite{Zeros}.
Increasing of coupling strength $\lambda$ results in grows of the anomalous pairing (\ref{barF}) and suppression of the initial quasiparticle peaks. 
In strong coupling the system shows divergent behavior at $\omega=0$. However, the divergence point might be addressed beyond the conformal limit \cite{Bagrets} $\omega \lesssim J/\left(N \log N\right)$. This changes the scaling of the SYK Green's function from $1/\sqrt{\omega}$ to $N\!\log N \sqrt{\omega}$ in the infrared. 

\begin{figure}[ht!!]
\center
\includegraphics[width=1.\linewidth]{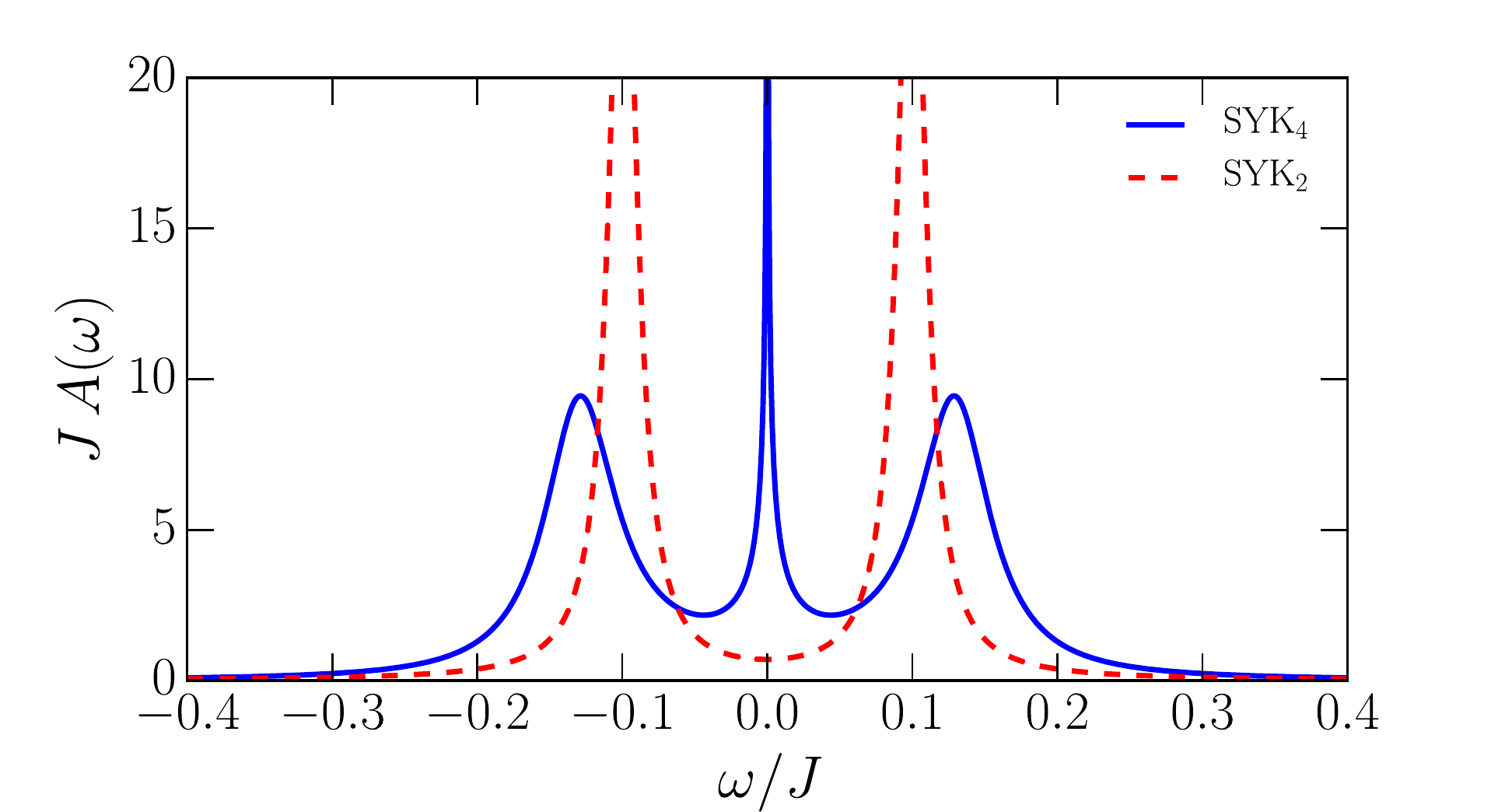}
\caption{\small \label{fig:LDOS24} \textbf{Density of states} in the quantum dot coupled to the \textbf{SYK/SYK$_2$} model at zero temperature. The coupling strength is $\lambda=0.1 J$ and the frequency of the single state is $\Omega_d=0.1 J$.}
\end{figure}

In FIG. \ref{fig:LDOS24} we show, that the behavior of the spectral function of the quantum dot coupled to the SYK model is qualitatively different from the SYK$_2$ case. The SYK$_2$ model, mentioned above, describes disordered Fermi liquid and has a constant density of states $\propto 1/J$ in the long time limit. 

\begin{figure}[ht!!]
\center
\includegraphics[width=1.\linewidth]{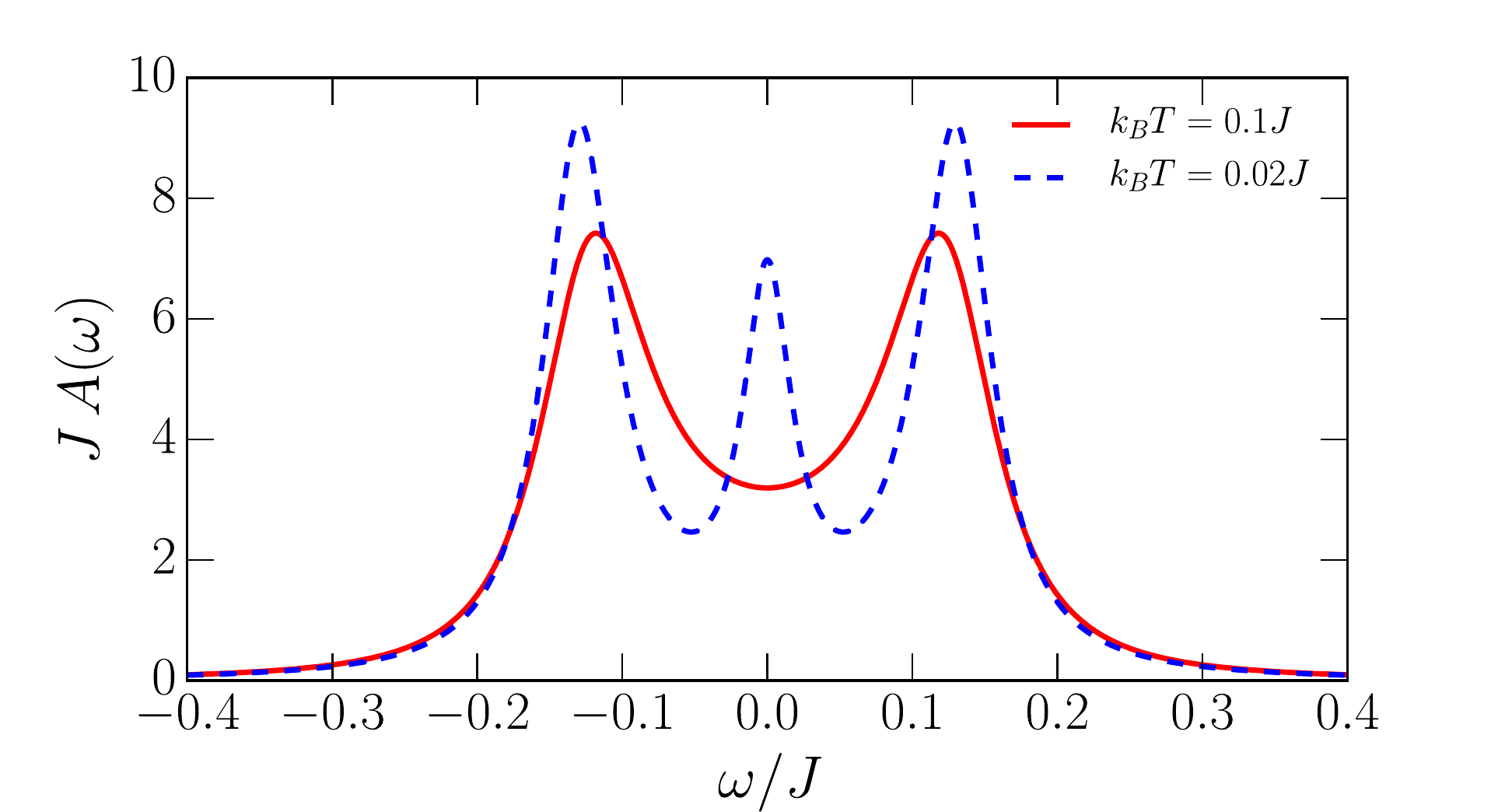}
\caption{\small \label{fig:LDOS_T} \textbf{Density of states in the quantum dot at finite temperature} as a function of frequency. The parameters are $\lambda/J=0.1=\Omega_d/J$.}
\end{figure} 

At finite temperature the saddle-point solution of the SYK model is given by \cite{Sachdev_BH}:
\begin{equation}
G^{R}_\gamma(\omega)= -\mathrm{i} \pi^{1/4}\sqrt{\frac{\beta}{2 \pi J}}\,  \frac{\Gamma\left(\frac{1}{4}-\mathrm{i}\frac{\beta \omega}{2 \pi} \right)}{\Gamma\left(\frac{3}{4}-\mathrm{i}\frac{\beta \omega}{2 \pi} \right)} \label{GR}, 
\end{equation}
where $\beta=1/T$ is inverse temperature and $\Gamma(x)$ is the Gamma function. We substitute the finite temperature SYK Green's function  (\ref{GR}) in the spectral function of $d$ fermion (\ref{LDOS}). 
FIG.\,\ref{fig:LDOS_T} demonstrates that the divergence around $\omega \sim 0$ in the quantum dot density of states is regularized at finite temperature. 

\textit{Conclusion ---}  In this paper, we have shown that a single-state spinless quantum dot becomes superconducting in proximity to a structure whose low-energy behavior can be captured by the Sachdev-Ye-Kitaev model. Anomalous averages are found exactly in the large $N$ limit and turn out to be odd functions of frequency. Appearance of non-zero superconducting pairing does not require any additional quantum numbers like spin, because it originates from non-locality of the SYK saddle-point solution. Induced superconductivity strikes in the intermediate coupling between the quantum dot and the SYK model. At stronger coupling, the quasiparticle peaks are smeared out on the background of the SYK quantum critical continuum. 
We propose to use the peculiar property of the induced gapless superconductivity in zero dimensions to characterize the solid-state systems, that can be described by the SYK model as an effective theory in a certain limit.

\textit{Acknowledgements ---} N.\,G. has benefited from discussions with Carlo Beenakker, Koenraad Schalm, Jakub Tworzyd\l{}o, Fabian Hassler, \.{I}nan\c{c} Adagideli, Sergei Mukhin, Alexander Krikun, Jimmy Hutasoit, Micha\l{} Pacholski, Andrei Pavlov, and  Yaroslav Herasymenko.
This research was supported by the Netherlands Organization for Scientific Research (NWO/OCW) and by an ERC Synergy Grant.

\begin{widetext}
\appendix

\section{Self-consistent derivation of the Gor'kov Green's function for the quantum dot variables}\label{app:SC}

The Euclidean action that corresponds to the Hamiltonian (\ref{H}) after disorder averaging  is
\begin{align} \nonumber
S=&{} \int_0^\beta d \tau \left[\bar{d}\left(\partial_\tau +\Omega_d\right) d + \frac{1}{2}\sum_{i=1}^N \gamma_i \partial_\tau \gamma_i\right] \\&{}-\int_0^\beta d \tau \int_0^\beta d \tau' \left[ \frac{\lambda^2}{2 N} \sum_{i=1}^N\gamma_i \left(\bar{d}-d\right)(\tau) \gamma_i \left(\bar{d}-d\right)(\tau') + \frac{J^2}{8N} \sum_{i,j,k,l=1}^N\gamma_i \gamma_j \gamma_k \gamma_l(\tau) \gamma_l \gamma_k \gamma_j \gamma_i(\tau') \right]  \,  \label{app:S}.
\end{align}
%Hereby we disregard spin-glass ordering \cite{SK}. 
We introduce four pairs of non-local fields as a resolution of unity:
\begin{align} \nonumber
1&{}=\int \mathcal{D} G_\gamma \, \delta\left(G_\gamma(\tau',\tau)-\frac{1}{N}\sum_{i=1}^N \gamma_i(\tau) \gamma_i(\tau')  \right)= \\ &{}=\int \mathcal{D} \Sigma_\gamma \int \mathcal{D} G_\gamma \exp\Bigg[ \frac{N}{2} \int_0^\beta d \tau \int_0^\beta d \tau' \, \Sigma_\gamma(\tau,\tau') \left(G_\gamma(\tau',\tau)-\frac{1}{N}\sum_{i=1}^N \gamma_i(\tau) \gamma_i(\tau')\right)\Bigg]\, \label{app:1_gamma}, \\ \nonumber
1&{}=\int \mathcal{D} G_d \, \delta\bigg(G_d(\tau',\tau)-\bar{d}(\tau) d(\tau')  \bigg)= \\ &{}=\int \mathcal{D} \Sigma_d \int \mathcal{D} G_d \exp\Bigg[\int_0^\beta d \tau \int_0^\beta d \tau' \, \Sigma_d(\tau,\tau') \bigg(G_d(\tau',\tau)-\bar{d}(\tau) d(\tau')\bigg)\Bigg]\, \label{app:1_d*d},  \\ \nonumber
1&{}=\int \mathcal{D} F_d \, \delta\bigg(F_d(\tau',\tau)-d(\tau) d(\tau') \bigg)= \\ &{}=\int \mathcal{D} \Xi_d \int \mathcal{D} F_d \exp\Bigg[\int_0^\beta d \tau \int_0^\beta d \tau' \, \Xi_d(\tau,\tau') \bigg(F_d(\tau',\tau)-d(\tau) d(\tau')\bigg)\Bigg] \, \label{app:1_dd}, \\ \nonumber
1&{}=\int \mathcal{D} \bar{F}_d \, \delta\bigg(\bar{F}_d(\tau',\tau)-\bar{d}(\tau') \bar{d}(\tau) \bigg)= \\ &{}=\int \mathcal{D} \bar{\Xi}_d \int \mathcal{D} \bar{F}_d \exp\Bigg[\int_0^\beta d \tau \int_0^\beta d \tau' \, \bar{\Xi}_d(\tau,\tau') \bigg(\bar{F}_d(\tau',\tau)-\bar{d}(\tau) \bar{d}(\tau')\bigg)\Bigg] \, \label{app:1_d*d*}.
\end{align} 
This allows us to rewrite the action (\ref{app:S}) as:
\begin{align} \nonumber
S=&{}\int_0^\beta d \tau \int_0^\beta d \tau' \bigg[\bar{d}(\tau)\bigg(\delta(\tau-\tau')\left(\partial_\tau +\Omega_d\right) +\Sigma_d(\tau,\tau')\bigg)d(\tau') +\bar{d}(\tau)\bar{\Xi}_d(\tau,\tau')\bar{d}(\tau') +d(\tau)\Xi_d(\tau,\tau')d(\tau') \\ \nonumber &{}+\frac{1}{2}\sum_{i=1}^N\gamma_i(\tau)\bigg(\delta(\tau-\tau')\partial_\tau +\Sigma_\gamma(\tau,\tau')\bigg)\gamma_i(\tau') - \Sigma_d(\tau,\tau')G_d(\tau',\tau)-\Xi_d(\tau,\tau')F_d(\tau',\tau)-\bar{\Xi}_d(\tau,\tau')\bar{F}_d(\tau',\tau) \\  &{}- \frac{N}{2}\left(\Sigma_\gamma(\tau,\tau')G_\gamma(\tau',\tau) +\frac{J^2}{4} G_\gamma(\tau,\tau')^4\right)-\frac{\lambda^2}{2} G_\gamma(\tau,\tau')\bigg(G_d(\tau,\tau')-G_d(\tau',\tau)+F_d(\tau',\tau) + \bar{F}_d(\tau',\tau)\bigg)\bigg]\, \label{app:S_nonlocal}.
\end{align}
Following Refs.\,\onlinecite{SY, Kitaev, Maldacena1, Sachdev_BH}, we assume that all non-local fields are odd functions of the time difference $\tau-\tau'$. In the large $N$, long time limit: $1 \ll J \tau \ll N$, self-consistent saddle-point equations  are
\begin{align}
\frac{\delta S}{\delta \Sigma_d}&{}=0 \Rightarrow G_d(\tau-\tau')=-\left\langle \mathcal{T}_\tau\, d(\tau) \bar{d}(\tau')\right\rangle \, \label{app:Gd_eq}, \\
\frac{\delta S}{\delta \Xi_d}&{}=0 \Rightarrow F_d(\tau-\tau')=-\left\langle \mathcal{T}_\tau\, d(\tau) d(\tau')\right\rangle \, , \;\;\; \frac{\delta S}{\delta \bar{\Xi}_d}=0 \Rightarrow \bar{F}_d(\tau-\tau')=-\left\langle \mathcal{T}_\tau\, \bar{d}(\tau) \bar{d}(\tau')\right\rangle \, \label{app:F_eq}, \\
\frac{\delta S}{\delta \Sigma_\gamma}&{}=0 \Rightarrow G_\gamma(\tau-\tau')=-\frac{1}{N}\sum_{i=1}^N\left\langle \mathcal{T}_\tau\, \gamma_i(\tau) \gamma_i(\tau')\right\rangle \Rightarrow G_\gamma(\mathrm{i}\omega_n)^{-1}=\mathrm{i} \omega_n -\Sigma_\gamma(\mathrm{i}\omega_n)\simeq -\Sigma_\gamma(\mathrm{i}\omega_n) \, \label{app:G_eq}
\end{align}
and
\begin{align}
\frac{\delta S}{\delta G_d}&{}=0 \Rightarrow \Sigma_d(\tau-\tau')=\lambda^2 G_\gamma(\tau-\tau') \, \label{app:Sd_eq}, \\
\frac{\delta S}{\delta F_d}&{}=0 \Rightarrow \Xi_d(\tau-\tau')=-\frac{\lambda^2}{2} G_\gamma(\tau-\tau') \, ,\;\;\;\frac{\delta S}{\delta \bar{F}_d}=0 \Rightarrow \bar{\Xi}_d(\tau-\tau')=-\frac{\lambda^2}{2} G_\gamma(\tau-\tau')  \, \label{app:Xi_eq}, \\
\frac{\delta S}{\delta G_\gamma}&{}=0 \Rightarrow \Sigma_\gamma(\tau-\tau')=J^2 G_\gamma(\tau-\tau')^3+\frac{\lambda^2}{N} \bigg(2 G_d(\tau-\tau') -F_d(\tau-\tau')-\bar{F}_d(\tau-\tau')\bigg) \simeq J^2 G_\gamma(\tau-\tau')^3\, \label{app:S_eq}.
\end{align}
Green's functions of the fermion in the dot enter the equation for the Majoranas self-energy (\ref{app:S_eq}) as $1/N$, so we neglect them in the large $N$ limit.
Thus, equations (\ref{app:G_eq}, \ref{app:S_eq}) are decoupled from the quantum dot and become the standard SYK Schwinger-Dyson equations \cite{Kitaev, Maldacena1}  $G_\gamma(\mathrm{i}\omega_n)^{-1}= -\Sigma_\gamma(\mathrm{i}\omega_n)$ and $\Sigma_\gamma(\tau)=J^2 G_\gamma(\tau)^3$ with a known low-frequency solution 
\begin{align}
G_\gamma(\mathrm{i}\omega_n)=-\mathrm{i} \pi^{1/4} \frac{\mathrm{sgn}(\omega_n)}{\sqrt{J |\omega_n|}} \label{app:G_SYK}
\end{align}
at zero temperature, where $\omega_n = \pi T(2n+1)$ are Matsubara frequencies.
Meanwhile, the bare SYK Green's function (\ref{app:G_SYK}) enters the self-energies of the quantum dot (\ref{app:Sd_eq}, \ref{app:Xi_eq}), that, according to the definitions (\ref{app:1_d*d}, \ref{app:1_dd}, \ref{app:1_d*d*}), give both normal ($\bar{d} d$) and anomalous ($\bar{d} \bar{d}$, $d d$) components of the effective action for the $d$ fermion.

The effective action for the fermion in the quantum dot is given by
\begin{align}
S=\frac{1}{2}\sum_{n=-\infty}^{+\infty} \begin{pmatrix}
\bar{d}_n & d_{-n}
\end{pmatrix} \begin{pmatrix}
-\mathrm{i} \omega_n + \Omega_d+\lambda^2 G_\gamma(\mathrm{i}\omega_n)& -\lambda^2 G_\gamma(\mathrm{i}\omega_n) \\ -\lambda^2 G_\gamma(\mathrm{i}\omega_n) & -\mathrm{i} \omega_n- \Omega_d+\lambda^2 G_\gamma(\mathrm{i}\omega_n)
\end{pmatrix} \begin{pmatrix}
d_n \\ \bar{d}_{-n}
\end{pmatrix} \, \label{app:S_eff},
\end{align}
so that the Gor'kov Green's function \cite{Gorkov} composed from (\ref{app:Gd_eq}, \ref{app:F_eq}) is found exactly in the large $N$ limit:
\begin{align}
\mathcal{G}(\mathrm{i} \omega_n)^{-1}=\begin{pmatrix}
\mathrm{i} \omega_n - \Omega_d-\lambda^2 G_\gamma(\mathrm{i}\omega_n)& \lambda^2 G_\gamma(\mathrm{i}\omega_n) \\ \lambda^2 G_\gamma(\mathrm{i}\omega_n) & \mathrm{i} \omega_n+ \Omega_d-\lambda^2 G_\gamma(\mathrm{i}\omega_n)
\end{pmatrix} \, \label{app:GG}.
\end{align}
The analytic continuation to the real frequencies $\mathrm{i} \omega_n \to \omega
+ \mathrm{i} \delta$ with $\delta=0^+$ gives the retarded Green's function in the particle-hole basis:
\begin{align}
\mathcal{G}^R(\omega)= \frac{1}{\left(\omega + \mathrm{i} \delta\right)\left(\omega + \mathrm{i} \delta -2 \lambda^2 G^R_\gamma(\omega)\right) -\Omega_d^2}\begin{pmatrix}
\omega + \mathrm{i} \delta + \Omega_d-\lambda^2 G^R_\gamma(\omega)& -\lambda^2 G^R_\gamma(\omega) \\ -\lambda^2 G^R_\gamma(\omega) & \omega + \mathrm{i} \delta- \Omega_d-\lambda^2 G^R_\gamma(\omega)
\end{pmatrix} \, \label{app:GGR},
\end{align}
where $G_\gamma^R(\omega)=-\mathrm{i} \pi^{1/4} \mathrm{e}^{\mathrm{i} \pi \mathrm{sgn}(\omega)/4} \left(J|\omega|\right)^{-1/2}$.

\end{widetext}

\end{document}